# Directors' duties in the age of agentic Artificial Intelligence: How can boards navigate corporate purpose and the stakeholder interests of employees around AI adoption?

## 1. Introduction

Artificial Intelligence is animating the fourth and fifth industrial revolutions at a speed of invention that is breathtaking. Presenting transformative opportunities for companies on a spectrum from low level automation to general purpose AI and agentic AI, boards are drawn towards AI in terms not just for the efficiencies it offers, but also how it promises entirely new business models and ways of operating. While actors such as Chief Information Officers and AI Governance and Ethics Committees loom large in corporate structures, consideration of questions around deployment or non-deployment of AI by the company falls squarely within directors' purview as part of boards' strategy mandate. Directors' duties complement other market-based controls on managerial conduct and indeed legal and ethical ones such as obligations under the EU AI Act.[1]

Functionally, Balkin asserts that "[p]eople will substitute robots and AI agents for … humans. But they will do so only in certain ways and only for certain purposes." (Balkin 2015, p.46). The human/object distinction that Balkin draws rings true if the substitution is partial, but the substitution may prove substantial and even complete as AI becomes more sophisticated and versatile. For example, with the right coding, agentic AI assistants are being set up to carry out software development tasks and to manage workflows and communicate and cooperate with humans in relation to the assigned tasks.

[1] Artificial Intelligence Act Regulation (EU) 2024/1689 of the European Parliament and of the Council of 13 June 2024 OJ L. 12.7.2024 ("the EU AI Act").

The duty on directors to act in the best interests of the company places a high level brake on boards' autonomy as they grapple with questions of strategy, opportunity and risk including around operationalising AI in their companies' operations. The fiduciary duty on directors to act in the best interests of the company, a venerable duty, often regarded as the most fundamental of directors' duties, acts as a curb on directors' agency in the exercise of discretionary power on behalf of the shareholder body, with the objective that directorial decision-making serves the interests of the company.[2] While a sub-field of management and business scholarship examines how AI can transform business productivity and profitability (Rubab, 2023, Filippucci et. al, 2024), by contrast, with a few exceptions, (Ahern, 2024; Brand, 2024) the implications of AI adoption for directors' duties remain under-studied.

An interesting paradox presents in this domain. AI is new, exciting and extremely fast-moving in terms of its ever-expanding technical capabilities and use cases. By contrast, directors' fiduciary duties including the best interests duty are long-established and slow moving in their development. That does not necessarily signify that they are not up to the task of governing how directors ought to behave in the age of AI. Indeed, as a broadly expressed, duty that is mandatory in application, context-agnostic and applied in all areas of corporate decision-making, the best interests duty is flexible enough to be able to respond to new contexts including AI even if it is rarely enforced.

[2] Key cases include *Hirsche v Sims* [1894] A.C. 654; *Re Smith and Fawcett Ltd* [1942] Ch. 304; *Brady v Brady* [1989] 1 A.C. 755.

The focus of this article is not on directors of the leading edge AI developer companies who put AI solutions to the market, but rather on companies who are actual or potential adopters of such AI systems. Directors of these companies want to ride the coattails of AI invention. Consideration of why, when and how to deploy AI in a company's business goes to the heart of their fiduciary duty to act in the best interests of the company. Directors do not want their companies to be left behind and lose market competitiveness. Yet AI adoption is not without risk. First, complex questions arise for directors around discerning what is in the best interests of the company. Secondly, the tricky conundrum of how to approach the potentially competing interests of employees emerges given that a board decision to adopt AI will often have knock-on implications for the company's workforce in terms of changes to work, altered work patterns and reduced headcount. The crux is that, in the great AI goodbye, boards are choosing to replace human workers with AI. As we enter further into the age of agentic AI that is autonomous, multimodal and dynamic (Hosseini and Seilani, 2025; Patel & Chauhan, 2025), this pattern is likely to deepen.

The question of how corporate purpose is defined matters for AI adoption, being closely coupled to any consideration of the best interests duty in the common law world. Yet there is an identifiable gap in the literature as regards the positioning of employees as corporate stakeholders within the context of AI adoption-related decisions of the board. This matters because directors may find themselves caught in a difficult place between, on the one hand, the urge to find competitive advantage in AI innovation, and on the other, awareness of the role continuity and welfare concerns of workers. Against that backdrop, this article poses the question of to what extent existing statutory formulations of the fiduciary best interests duty allow for employee interests to be considered by directors in making decisions around AI adoption. In terms of method, theoretical, doctrinal and real world insights are blended. Examining AI through a relatively wide lens including comparative study of the best interests duty, soft law, and a law in context approach that highlights the power of real world nudges allows for novel, real world insights contributing to corporate law scholarship on directors' duties in the era of AI and filling knowledge gaps in the field.

Section 2 provides a brief background context around what AI adoption offers to deployer companies and the resulting changing nature of work. Section 3 introduces corporate purpose and the classical alignment of the interests of the company with shareholder primacy and profit maximisation. The recognition of the interests of employees as stakeholders in the corporate endeavour is then contextualised to the interface with AI and its impacts on the workforce. In Section 4 the intriguing question of whether in an age of AI ascendancy AI may warrant being given stakeholder status is indulged. From that springboard, the article provides a comparative doctrinal analysis of the nuances of four core models of codified best interests duty in the common law world with the object of excavating the parameters which they provide directors to balance employee interests against shareholder interests in AI-driven profit maximisation. A continuum of four different codified corporate purpose models, namely, shareholder primacy, enlightened shareholder value, stakeholder friendly and stakeholder value are examined. Section 5 begins by considering the typical shareholder primacy doctrinal model and how it governs directors decision-making around AI adoption decisions in a manner that is welfare-enhancing for shareholders. It also highlights cases where directors' judgment may be second-guessed, including where they authorise 'AI washing'. This provides a solid foundation for an analysis in Section 6 that focuses on providing a determination of the impact of differential legislative recognition of stakeholder interests within individual best interests' duty statutory formulations on the ability of directors to take on board employee interests in strategic AI decision-making. In moving from stakeholder theory to practice, of particular interest for corporate law scholars are the shifts around stakeholderism in codifications of the best interests duty in the United Kingdom, Canada, New Zealand and India that provide pathways for directors to more inclusive forms of corporate purpose. Section 7 highlights the restricted pathways open to employees who wish to allege breach of the best interests duty. The law here is ill-fitting as a means of redress for AI adoption that impacts or displaces human employees. The exceptionally limited ability of employees to operationalise the best interests duty through enforcement suggests the need to cast the net wider than the catchment of the best interests duty in understanding how their interests may receive direct or indirect support. This is multi-faceted territory. The Section goes on to examine the efficacy of the contribution made by some soft law corporate governance precepts to employee engagement and voice.

A central premise of this article is that a strict shareholder primacy lens is shortsighted. Directors should appreciate that the ripples of the best interests duty and the stakeholder debate extend far beyond both the wording and its enforcement trajectory. In the real world how boards handle interactions with employees matters. Accordingly, Section 8 advocates for a law in context approach, arguing that the content for the contextual application of directors' duties around AI adoption is supplied not just by directly enforceable norms that emerge from case law, but also is significantly influenced by a wider accumulation of norms including those deriving from emerging societal expectations. Directors' decision-making and communications around planned employee impacts as a consequence of AI integration present clear potential for negative spillover reputational and goodwill implications. As such, a law in context approach is appropriate to the need to consider other sources of meaning and content for acting in "the best interests of the company" outside of statute and judicial ruling. In that light, AI adoption should not be tone deaf to the tensions that it brings for employees. Boards of well-motivated companies can choose to go beyond a minimum compliance approach and to meaningfully consider the interests of their existing employees and their future work prospects through transparency, employee engagement and, where possible, providing employee retraining to reskill employees.

## 2. AI Adoption, Companies and Labour Market Disruption

AI is a disruptive technology lauded by the business world for delivering efficiency and productivity gains that aid the bottom line. On the other hand, the technology is still evolving and the level of AI-readiness in companies is heterogenous, varying hugely across company size and industry. Like companies, AI is far from monolithic and there is a vast array of ways in which companies can use Generative AI to drive automation and augmentation (Holmström & Carroll, 2025) from simple efficiencies to business model transformation. Companies' adoption of AI thus occurs on a vast continuum from the productivity gateway of Generative AI digital tools such as Co-Pilot (assisted by its integration into Microsoft Office), ChatGPT and Gemini for summarising online meetings and email writing to extensive use of agentic AI systems for coding, automation of business operations and manufacturing processes.

Before the arrival of Generative AI, many companies that had not been proactive in identifying a meaningful internal use for AI. However, the arrival of Generative AI has been game-changing for many companies. Companies are using it for content generation for marketing and triaging customer service queries. As regards internal AI adoption in the boardroom itself, things have moved on from scholarly speculation about its potential use of AI (Locke & Bird, 2020) to directors turning to AI tools to help them to deal with the paper mountain that the board papers present including through using Generative AI to formulate constructive challenge questions for board meetings (Shekshnia &Yakubovich, 2025). While many companies are making ready use of AI for process automation and Generative AI tools which are cost-effective to deploy, using AI in a transformative way tends to be more common for the largest companies who have the resources and the know-how to make the investment. On the other hand, there is some evidence that medium sized companies can be more nimble than large behemoths such as the Big Four accountancy firms at embedding AI into the core of their work (Kissin, 2025a). It is also worth acknowledging that thus far AI has most impact on knowledge industries and AI being used in the production of goods and services still remains the exception rather than the norm (Reiter, 2025). It is, however, very effective in optimising supply chain management.

There is a burgeoning scholarship across the disciplines on AI-related labour market disruption and the future of work (Frey & Osborne, 2023; Kraus et al., 2023). The disruptive effects on the labour force are huge and growing. AI is creating new career opportunities such as AI ethics and governance specialists (Passador, 2025). However, in the great AI goodbye boards are choosing to replace human workers with AI in a bid to become more agile and leaner while boosting profitability. This has seen a crunch for software engineers, but also for those involved in data entry, manufacturing, logistics, even HR and marketing. Companies are moving en masse to using chatbots to cut customer services support costs while AI agents and robots are leading to household names laying off thousands of workers.

The speed at which this is happening is radically shaking up labour market path dependencies. Amid the inevitable job displacement that AI brings, higher chain jobs are often safer than lower chain ones than can be automated. The Chief Executive of Anthropic has predicted that 50 percent of entry-level white collar jobs may be wiped out within the next five years (VandeHei & Allen, 2025). For enterprise AI adoption, depending on the company and the industry, automation and AI-related employee displacement is a short or longer game. AI adoption fundamentally alters the contribution of the employee and in some cases their contribution is being written out of the equation as AI can do it faster and without payment. Some employees are lessening their own value proposition by over-delegating their work to AI tools with poor results (Niederhoffer et al., 2025). As the nature of work shifts, to retain value, the role of the employee of the future is likely to increasingly become that of skilled AI prompt engineer, trainer and overseer rather than doer. As AI's benefits to companies are pitted against job displacement, in Section 3 we probe what lessons corporate governance theory yields in relation to how boards should recognise shareholder interests and treat employees as stakeholders in the firm.

The point in the trajectory of AI adoption matters to identifying who are the company's stakeholders and defining what are their interests. Whether the interests of employees are looked at by directors in the short, medium or the long-term leads to differences of approach. Embedding AI successfully requires real cultural change and organisational buy-in (Li, Zhu & Hua, 2025). Amid talk about the future of work, it is worth bearing in mind that in some companies, industries and countries substantial AI-related adaptions have already occurred and bedded down. For instance, in Japan it is commonplace for human employees to work alongside robots who also provide an answer to the labour force challenge of an ageing population. In addition, adopting a long-term lens, some human roles and functions are likely to have faded out of the picture in a given company, and with them the interests of the former cohort of workers who previously carried them out (or at least they are now identified with their new status quo which may or may not be as an employee of the company).

**3. Corporate Purpose, Employees and AI Adoption**

Understanding the application of the best interests duty in this sphere requires conceptualising the lens through which interests are viewed (i) as a matter of corporate governance theory, and (ii) in doctrinal practice. This has significance where the prognosis that AI can be relied on to add value, increase profits and rally share prices, is juxtaposed alongside rapid obsolescence of employee roles.

Corporate purpose is generally deeply connected to the generation of profit.[3] Shareholders as residual owners delegate the day to day running of the company to the directors. On a shareholder primacy understanding, directors' duty is to manage the company in the interests of the shareholders (Berle, 1931; Friedman, 1970; Hansmann and Kraakman, 2001). As they are investors and residual owners, they have primacy and acting in the company's interests is associated with the interests of the shareholders in profit maximisation and capital appreciation. When a board makes decision to adopt a strategy to drive automation and AI systems within a company which is designed to reduce costs and increase organisational efficiency and supply chain management with positive impacts for financial performance, that is a shareholder-aligned decision within the dominant shareholder primacy corporate purpose model.

---

[3] B-corps/social enterprises are outside the scope of this paper. On the potential for broader pro-social purposes to be adopted by B-corps see Klettner, 2024.

Indeed, directors have been said to have "strong incentives to prioritize shareholder interests and limited incentives to prioritize other groups" (Hutchison, 2025, p.35). A company within the Milton Friedman mould (Friedman, 1970) may simply be driven by profit and efficiencies and the directors may exhibit no particular conscience or willingness to go beyond the letter of the law to engage in purely ethical behaviour for the benefit of other stakeholders. That said, it would be wrong not to acknowledge that more accommodating and inclusive concepts of value creation including stakeholder capitalism have made a discernible impact, recognising that is possible to combine profit with concern for stakeholders (Business Roundtable, 2019; Edmans, 2020, Mayer, 2018).

Stakeholder theory has both descriptive and normative dimensions (Donaldson & Preston, 1995) and within it stakeholder interests have been well explored in the context of corporate law and corporate governance scholarship, (Blair & Stout, 1999; Dodd, 1932; Fairfax, 2005; Freeman, 1984; Marshall & Ramsay, 2012). However, the specific context of the workforce as a stakeholder in AI-related decisions of directors suffers from a lack of specific scholarly consideration. Employees qualify as corporate stakeholders by virtue of their ability to affect, and be affected by, the corporation (Donaldson & Preston, 1995; Freeman, 1984). Team production theory emphasises the role of a team of actors in the company's outputs and success with directors acting as “mediating hierarchs whose job is to balance team members' competing interests in a fashion that keeps everyone happy enough that the productive coalition stays together” (Blair & Stout, 1999, p.281). In addition, stakeholder theory emphasises ethical capitalism that values those who have a stake in value creation and the idea of sustainable value creation. Furthermore, while the employee headcount may reduce in a company to reflect AI adoption, ensuring the welfare of those employees retained could prove mission-critical to ensuring that AI strays on track as they become key assets under team production theory.

In any company where AI investment is more than rudimentary, there is likely to be a conflict between viewing this through a reductive shareholder primacy optic that values AI-driven efficiencies and cost-cutting in service of profit and how it will be viewed through the stakeholders lens as regards impacts on the employee constituency in the short, medium and long-term. Organisational culture and technology-driven changes affect dynamics around how employees work and interact and perceive their place and respect for the company. Employees rely on employment, not only for livelihood and security, but also for a sense of purpose and self-actualisation (Maslow, 1998). Employees may nevertheless be regarded by companies as disposable in the age of AI. Shareholder primacy interests in optimising a company's business processes to use AI to the benefit of the bottom line butt up against those of employees. AI integration can lead to shedding of the workforce to add productivity gains and reduce headcount costs or it may occur by stealth in the labour market to affect future employment prospects in the form of "low hire, low fire". AI has also unleashed increased precarity in work conditions amid a "wait and see" approach by companies to how much AI could do in place of employees. When the value proposition of AI does become apparent, automation "offers the ultimate exit from the costs and risks associated with human labour" (Estlund, 2018, p. 254). With huge and seemingly permanent cost savings to be made, boards may determine that the company can afford to write once-off redundancy cheques to workers carrying out repetitive tasks on the back of ushering in cost-effective AI 'workers' and systems.

Of course through the strict view of shareholder primacy, the interests of such employees are legal not moral, protected only outside directors' duties and company law by regulatory and contractual methods (Hansmann & Kraakman, 2001). Morality will have no purchase, and legally provided the processes for selection for redundancy are fair, the fact of replacement by an AI tool will not of itself be a ground for unfair dismissal.[4] Moving to a stakeholder theory perspective, engaging transparently and in good faith with employees around the organisation's future of work and listening to their concerns is valuable given employees' interests in actual and potential AI-driven impacts on their role and career path. Planned and potential changes to work patterns and threats to jobs security are of concern to the workforce. As discussed in Sections 7 and 8, process needs constitute the need for employees and their representatives for transparency and being heard in terms of AI-driven impacts within the company and fairness in terms of how they are treated as employees around AI-adoption and adaptation.

[4] Workplace Relations Commission case (ADJ-00054525) in Ireland relating to a TikTok employee whose monitoring role was to be replaced by an AI tool.

Both lower order and higher order employee contributions to firm outputs may be altered, diminished and rendered superfluous as AI is increasingly integrated into firms. A question that arises is whether the stake of some employees is potentially diminishable or extinguishable in the face of the decreasing relevance of those employees' input to the future of the firm? Or normatively, ought the burden on the board to "do right by those employees" increase rather than decrease? A shareholder primacy lens would treat this as a simple matter of employment law. As discussed in Section 8 below, a stakeholder-inclusive lens would take a more far-seeing, ethical approach that asks the board to go further to consider how best to meet the needs of those employees. Having examined corporate governance theory-based understandings of the interests of shareholders and employees, in Section 5 the best interests duty is unpacked through the lens of shareholder primacy formulations. First, we turn briefly to the possibility of AI being recognised as a stakeholder.

**4. AI as a Stakeholder in an Age of AI Ascendancy?**

Ought there to be a need to potentially balance the needs of employees with those of AI agents? AI is non-sentient and to date has not been granted legal personhood. However, as the substitutability of AI in the workplace erodes human workers, a question to ponder is whether AI agents could deserve recognition as a stakeholder in the corporate purpose debate by virtue of its functional quasi-worker status.

The debate around stakeholder theory can be traced back almost a century to Dodd (Dodd, 1932). Subsequently, Freeman's influential work supplied the definition that a stakeholder is “any group or individual who can affect or is affected by the achievement of the organization’s objectives” (Freeman, 1984, p.46).

Significantly, corporate law and scholarship has not limited admission to stakeholder categories to human/sentient stakeholders. The environment has been recognised as a stakeholder as seen in the non-exhaustive list in s.172((1) of the Companies Act 2006 (Rosel, 2018). Drawing succour from this, in applying Freeman’s definition of a stakeholder, it is not difficult to see AI as both affecting and being affected by a company’s objectives. How AI systems, particularly generative AI tools, are deployed, including how they are programmed, trained on data sets and used by a company will involve an ongoing interaction that dynamically impacts not only the intrinsic nature and capabilities of the AI tool during its lifecycle, but also the company where it is deployed and, more broadly, on society. However, corporate law is not well placed to specifically guide directors on AI’s impacts. Instead, regulatory and governance frameworks on AI such as the AI Act and industry standards are needed to ensure that AI is safe, trustworthy and human-centric and meets the values of our society. This is complex and remains an evolving space.

Team production theory would be the most apt theoretical hook to justify AI as a stakeholder as AI agents become indispensable to achieving business ends. That begs the question of the necessity for an AI agent to have legal status as a prerequisite to being recognised as a stakeholder, and also what its “interests” might identified to be. In terms of standing, lack of a legal status has not proved critical to recognition as a stakeholder as the gradual recognition of the environment as a non-human stakeholder in corporate law systems has shown (Phillips & Reichart, 2000). While discussion around granting legal status to AI tends to centre around accountability and liability concerns, looked at simply as a compliant functional worker, it is easier to conceive of AI as a putative stakeholder. Indeed, a fascinating developing line of scholarship engages with the moral and legal status of humanoid robots and their potential rights (Gunkel, 2018; Schröder, 2021; Silva Sampaio, 2024). This also brings in the developing discussion of robo-ethics which could provide a basis for an inclusive moral stakeholder approach that recognises AI agents as having interests that go beyond those with a basis in law. For instance, Silva Sampaio argues that using an interest theory could open up wider possibilities around robots being holders of passive rights (rather than the ascribed enforceable rights of humans) (Silva Sampaio, 2024). These ideas of course remain speculative.

What is clear is that as AI's unfolding autonomy and agency escalate, rendering permanent the displacement of certain roles and functions of human workers, we enter a new era where conceptually AI agency and interests as tied to corporate purposes ascend into the realm of relevance. It is predicated that this point will be reached at the paradigm shift when advancements in machine learning and deep neural networks enable AI to summit the much-anticipated Artificial General Intelligence phase to marshal cognitive abilities that are equivalent to or surpass those of their human masters. This point presents a threshold where it would be appropriate to consider AI as a potential stakeholder, conceptually, if not legally. But perhaps we do not need do wait until then given the arrival of agentic AI - LLM tools that can autonomously act often as part of a system, aiding productivity by carrying out a wide variety of tasks and workflows with limited to no human supervision.

A detailed exploration of the nature of AI's stake in the corporation lies beyond the scope of this article. However, speculatively, accepting that the constituencies of stakeholders are not fixed and that AI on its face meets the definition of a stakeholder in that affects and is affected by the company endeavour, opens the potential for further normative claims. Then, proponents of team production theory could argue that, alongside other team members, AI systems too should accrue the benefit of Blair and Stout's normative claim that "directors should be viewed as disinterested trustees charged with faithfully representing the interests not just of shareholders, but of all team members," (Blair & Stout, 1999, p.286). What that would look like is difficult to say. In any event, the problem of competing interests of stakeholders in the corporation remains unresolved.

It is not difficult to see the potential for a head-on clash here between the interests of employees versus those of AI that may be considered to align with those of shareholders in efficiency and profit maximisation. It is posited that the stakeholder interests of an AI agent would be to be in service of the company - carrying out its assigned tasks as effectively and efficiently as possible to achieve its goals effectively. An appropriate degree of human interaction and oversight and regular technical maintenance and algorithmic auditing would respect those interests and align with the duty on directors to act in the best interests of the company.[5] Thus there should be significant goal alignment rather than goal conflict between the interests of AI as a stakeholder and the shareholder primacy goals of the directors.[6]

[5] More broadly, it is also vital for boards to carry out their strategy and monitoring roles to ensure that AI agents remain "on task" and do not independently act in a way that is harmful to the company's interests and those of the wider public.

## 5. Serving Shareholder Interests around AI Adoption within the Base Shareholder Primacy Model

The duty to act in the best interests of the company is a duty traditionally exercised "for the benefit of the company as a whole".[7] Foundational statutory codifications of directors' duties in jurisdictions such as Australia,[8] Ireland[9] and South Africa[10] prioritise neoliberalist conceptions of capitalism founded on shareholder primacy. As jurisdictions which have embraced the dominant shareholder primacy model, broad autonomy within defined parameters is available to directors of companies in these jurisdictions in their discretionary decision-making. There are differences of emphasis according to whether the interest of the company are defined as the short-term interests of shareholders or the long-term interests of the company as an entity.[11] While a short-term approach may be associated with undue risk-taking in search of short-term share price gain, on the other hand, a long-term value creation approach indicates an emphasis on survival of the company for the long-term (Keay, 2008).

---

[6] We leave out of consideration AI agents that fail to act for proper purposes and go rogue. For illustration see HAL's Stanley Kubrick's science fiction film 2001: A *Space Odyssey*.

[7] *Allen v Gold Reefs of West Africa Ltd* [1900] 1 Ch. 656, 671.

[8] Corporations Act 2001, s.181(1)(a).

[9] Companies Act, 2014, s.228(1)(a).

[10] Companies Act 71 of 2008, s.76(3)(b).

[11] *Gaiman v National Association for Mental Health* [1971] Ch. 317 (looking at the interests of present and future members).

AI is shaking up market practice. Indeed, asset managers are beginning to dispense with the tried and trusted route of reliance on external proxy advisory firms to analyse shareholder proposals in favour of the analysis provided by internal AI tools (Adams et al., 2026). There is a potential concern that the availability of AI tools to boards and investors to analyse market data in real-time could contribute to the type of myopic, short-termism that can also be driven by ad hoc and quarterly reporting (Landy, 2025). The ability they provide to rapidly parse and analyse corporate disclosures and decisions, fear of real-time AI-informed investor decision-making by could contribute to a heightened short-term lens being adopted by boards (Jing, 2025). This feeds into a broader debate on whether benefits from increased reporting frequency in terms of increased market efficiency are outweighed by potential incentivisation of a short-term corporate outlook (Ernstberger et al, 2017). As “AI augmented investors” (Lee, Oh & Wang, 2026, p.1) use advanced AI analytics to inform algorithmic trading, AI-driven sentiment analysis that informs share price in close to real-time could further constrain boards to stay within the safe parameters of shareholder primacy norms rather than attempting to incorporate broader stakeholder values. The use of an AI tool by Deutsche Bank to gain the edge in rapid buy-and-sell trading response to the release of UK’s budget shows this type of mentality in action (Partington, 2025). On the other hand, increased market transparency makes markets more efficient. Furthermore, AI models may protect creditors as models using real-time pattern recognition can discern signals of potential for liquidity risk/financial distress that may not otherwise be immediately obvious to a human (Ahern, 2026).

Operating through a shareholder primacy lens, in acting in the company's interests, directors ought to be alert to how AI could deliver competitive edge for the company. More broadly, it is also vital for boards to carry out their strategy and monitoring roles to ensure that AI agents remain "on task" and do not independently act in a way that is harmful to the company's interests and those of the wider public. As well as maximising revenue, levelling up AI systems can radically transform business processes. AI-wrought efficiencies can shave firm costs by with AI-modifications to operational processes from manufacturing to customer services create time and labour cost efficiencies. That said, a decision by a board on whether and how much to move ahead with AI adoption at any given time is far from a simple one. Leaving aside adoption of the low hanging fruit of basic mass-marketed generic AI tools, boards have far more exploration and due diligence to do in terms of assessing the case for AI adoption for business processes that would require adaption for their business and considerable investment. Furthermore, while associated with delivering efficiency gains, AI's impact is far from universally positive given ethical issues and the potential for harm to arise (Bar-Gill & Sunstein, 2025; Smuha, 2021). This points up the need for nuanced assessments by directors around the duty to act in the company's best interests. While the duty is not prescriptive, in practice, compliance with the fiduciary duty to act in the company's interests should involve the directors considering how the advent of AI could impact the company in terms of both opportunities and risks that could present for the company's interests. Failure by the directors to steer the company to engage in strategic repositioning to take account of AI advancements could lead to loss of competitive advantage. Accordingly, as the capabilities of AI technology advance at a staggering pace, boards should regularly review the company's business model and strategy to determine its sustainability for the future and to make adjustments as needed to steer the ship forward in the company's interests (Climent et al., 2024).

While courts are traditionally loath to trespass on directorial decision-making by retrospectively second-guessing it,[12] a useful lever comes from soft law principles that floodlight how companies ought to proceed as a matter of best practice in approaching AI and technological change through encouraging boards to create a corporate culture that adapts to innovation. The UK Corporate Governance Code 2024 Guidance (Financial Reporting Council, (UK), 2025, para. 9, 14) refers to the importance of boards ensuring that management is on top of technological change and the use by the company of AI, for example, in reporting. In South Africa, boards subject to the King Code are expressly prompted to be strategically proactive around leveraging new technologies and their benefits for the company in an ethical and compliant fashion (Institute of Directors of South Africa, 2025, Principle 10, Recommended Practice 105). On the other hand, the corporate landscape is not heterogenous. Given the vastly differing nature, size and resources of companies AI penetration may be more or less relevant. Not surprisingly, amid AI hype, some companies are struggling to see how embedding AI technologies will deliver business value (Enholm et al. 2022). The value-add of AI may sometimes not justify the financial and organisational commitments.

Directors equally need to be aware of legal and ethical dimensions and other risks of harm requiring appropriate management and governance policies to be put in place. Boards need to be particularly alert to reputational issues that may arise from AI slop

---

[12] *Re Smith and Fawcett Ltd* [1942] Ch. 304.

whereby poor judgment when using Generative AI tools leads to shoddy work outputs including Generative AI hallucinated inaccuracies (Kissin, 2025b). Employees, contractors and suppliers may be using and interacting with AI (including through employees using it in a non-authorised fashion). As such, irrespective of company size and business model, a board decision to put in place a corporate AI governance policy is sensible.

### *5.1 Insulation of Directors' Decision-making*

Litigation on the best interests duty is extraordinarily rare as the test for breach and the rules on derivative claims act as a procedural sieve allowing only a very small subset of claims to pass through (Ahern, 2011; Akanmidu, 2018). Boards have breathing space to arrive at decisions around AI adoption knowing that they are pretty watertight from challenge from the perspective of the best interests' duty. Arguably, this protective approach is helpful to directors given that commercial decision-making around AI adoption is complex in a rapidly evolving field where there may be market penalties associated with going in too early or waiting too long, requiring multiple factors from potential benefits to risks to be weighed in the balance. This resonates with director primacy theory which emphasises the pre-eminence of directorial judgment and its insulation (Bainbridge, 2002).

AI deployment is quintessentially a matter for corporate strategy within the sphere of the board. With the duty to act in the best interests of the company not being prescriptive in relation to how it is be fulfilled, what it means to act in the company's interests or best interests falls to be defined by the directors and decisions and actions of the directors around AI adoption are well-insulated from being judged in breach of the duty to act in the company's interests in the event of a directors' duties challenge being mounted by shareholders where dissent arises around technology-driven change (Lehdonvirta & Ricucci, 2023). As regards shareholder dissatisfaction with AI adoption or non-adoption, irrespective of whether a shareholder primacy jurisdiction tests compliance with the duty in an objective fashion (Australia) or a subjective fashion (Ireland),[13] it is an uphill battle to first gain access to the courts and second to demonstrate that there has been a breach of duty. Judges always emphasise that they are not in the business of being entrepreneurial. Moreover, a subjective framing of compliance is highly deferential to directors' belief that they were advancing the company's interests leaving very little opening for a finding of breach under this duty.[14]

### *5.2 When Might a Challenge to Directors' Decision-Making Be Possible?*

There are a few scenarios where a shareholder challenge around AI-related decision-making could be well-founded in a shareholder primacy duty model. For countries with a subjective approach to testing compliance, doctrinal developments suggest that an objective approach is only warranted in two very limited and exceptional circumstances. The first of these would be where there has been a complete failure by the director to address their mind to the interests of the company before making a decision related to AI adoption - the *Charterbridge* test then applies.[15] The second circumstance where the subjective approach can be displaced is where there have been bad faith considerations or a director withholds material information from the board.[16] For instance, if a director chairing the board's technology committee did not apprise the board of material information in relation to known instances of problems with an AI system revealed in an audit, this would have the consequence that the board was acting on incomplete information in making a decision to approve further roll-out the system within the company's wider operations following a limited divisional trial. The director in such a case could be considered not to be acting in good faith.

[13] Corporations Act 2001, s.181(1)(a) (Australia); Companies Act 2014, s.228(1)(a) (Ireland).

[14] *Re Smith v Fawcett Ltd* [1942] 1All ER 542; *Regentcrest plc v Cohen* [2001] BCC 337; *Dynamo Recoveries Ltd v Nix* [2024] EWHC 3116 (Comm).

[15] *Charterbridge Corporation Ltd v Lloyds Bank Ltd* [1970] Ch. 62. See also *Extrasure Travel Insurances Ltd v Scattergood* [2003] 1 BCLC 598, [138]; *Re Debut Homes (in liquidation; Madsen-Ries v Cooper* [2020] NZSC 100.

One wonders if the unwillingness of the board of a company to seek out advice on the game-changing nature of AI for their business could be so egregious as to amount to a breach of the best interests duty. For instance, one would need to be stuck under a rock not to realise that AI is here and the failure to adapt the business model may lead it to being not just less competitive but even on a rapid path to obsolescence. However, realistically, if there is no bad faith element, the path of least resistance would be to pursue omissions of this kind under the independent duty to exercise care, skill and diligence where an objective approach of the reasonable director often complements a subjective one (as seen in s.174(2) of the Companies Act 2006 (UK). Indeed, case law supports the notion that a director should upskill in order to be able to appropriately undertake their role.[17]

However, if a decision made around AI adoption was so manifestly not in the company's interest – where it could badly affect the company's reputation – then it could not be taken to be a good faith decision. The *Dynamo Recovery* case that arose out of the

[16] *Re Spring Media Investments Ltd*; *Saxon Woods Investments Ltd v Costa* [2025[ EWCA Civ 708.

[17] *Australian Securities and Investments Commission v Healey* [2011] FCA 71; (Ahern, 2024).

Cambridge Analytica scandal highlighted that data analytics should not be used for unlawful purposes and that a decision to act illegally is clearly not in the company's interests.[18] It is not difficult to extrapolate from this that directors must not allow AI tools to be used for unethical or unlawful purposes such as surveillance.

AI is being presented as such a versatile contributor to the bottom line - an enabler of cost cutting, productivity gains and new products and service markets mean that companies do not want to look like they are being left behind. While there is a perception that AI is ubiquitous and that companies should look like they are on board that train, there nevertheless can be a gap between the promise of AI and what it is currently actually capable of delivering (Kim, 2026). Moreover, companies may be purveyors of hot air or "hype" around their adoption and integration of AI that does not match reality. "Hype" here is used in the sense of a company's promotion of its supposed adoption of AI in a way that may falsify or exaggerate the actual level of engagement and/or benefit being delivered. For example, Meta was regarded as fudging Lama 4's benchmarks to "puff up its performance" (Kim, 2026). However, as seen with the phenomenon of greenwashing, puffery of any kind can backfire on boards and, when uncovered, negatively impact market reputation and share price (Xu et al., 2025).

Furthermore, similar to greenwashing, AI washing through falsely claiming that the company's business model is AI-driven or exaggerating its impact may be considered unethical and sometimes illegal (Ozturkcan & Bozdağ, 2025). Reputational damage is a likely consequence of AI-washing being discovered so it is clearly not considered responsible behaviour (Selby, 2024). It is highly unlikely that a director's decision that the company should engage in AI washing in its public documents - pretending that AI is powering the business model or exaggerating it - would be considered a good faith and honest decision in the company's best interest.

[18] See the Cambridge Analytica case *Dynamo Recoveries Ltd v Nix* [2024] EWHC 3116 (Comm) in relation to political campaign tactics to discredit political opponents.

Financial reporting and securities laws buttress this conclusion. In the United Kingdom, s.90 of the Financial Services and Markets Act 2000, which enables a shareholder claim for loss arising from false or misleading statements in listing particulars, potentially provides a more direct route to remedy than a directors' duty claim. The US landscape has already yielded class actions featuring AI washing claims where it is alleged that corporations falsely claimed they were using cutting edge AI technologies had overstated their capabilities.[19] In the case of Applovin Corporation, the company suffered somewhat of a fall from grace amid SEC and shareholder scrutiny. A plethora of class action securities lawsuits[20] arose in circumstances where it is alleged that the corporation misled investors in relation to how an upgrade its digital ad platform, AXON 2.0, would lead to improvements powered by cutting-edge AI and that this had positively impacted ad metrics and revenue growth. To the contrary, it is alleged that this was a smokescreen and that the dubious practice of backdoor installations of the app onto user phones, amongst other factors was driving the company's mobile gaming and e-commerce penetration. Broader questions have been raised as to whether the board prioritised short-term gain over ethical compliance. The outcome of such lawsuits remains to be seen but in the meantime there is a consciousness that misleading information that talks up a company's AI-value proposition can have a correlated distortive effect on share price.

[19] Class actions are being taken on behalf of investors in Applovin Corporation and Skyworks Solutions.

[20] See, for example, *Quiero v Applovin Corp., et al.,* No.4:25-cv-00294 (N.D. Cal.) which alleges violations of s.10(b) and s.20(a) of the Securities Exchange Act of 1934.

The punitive approach of relevant sectoral regulators to AI washing takes it from morally reprehensible to legally problematic. The Federal Trade Commission ("FTC") has initiated proceedings against companies in relation to allegedly deceptive claims where AI-related tools were marketed while misleadingly promising unrealistic returns (Federal Trade Commission, 2025. See also Federal Trade Commission, 2024). Nor would it be considered wise for a director or board to sanction AI capabilities being overstated in a company's financial and annual reporting. The Securities and Exchange Commission ("SEC") has investigated instances of AI washing in the context of financial reporting and has charged investment advisers who misled investors about their purported use of new technologies in their business (Securities and Exchange Commission, 2024). In a cautionary tale, an investment advisory firm was fined by the SEC for falsely purporting to use AI to make investment decisions) (Wigglesworth, 2024). The ongoing *SEC* v. *Sangier*[21] case should also bring pause for thought. There is a joint Department of Justice and SEC case against the founder of Nate Inc. for AI washing for securities fraud in relation to €42 million in capital raised on the back of misrepresentations in the form of false and misleading statements to investors around the company's purported use of AI. While marketed as a leading edge highly automated shopping app powered by AI, machine learning and neural networks, it is alleged that in fact transactions were being manually processed by contract workers at the back end.

[21] *SEC v. Saniger,* No. 25-cv-02937 (S.D.N.Y. filed Apr. 9, 2025); *United States v. Saniger*, No, 25-cr-00157 (S.D.N.Y.).

What these developments show is that directors should not be complacent about decisions made to engage in hype around their company's use of AI. This Section has explored how shareholder interests are met within best interests duty formulations founded on a shareholder primacy conception of corporate purpose. In the next Section we consider how the four different models of corporate purpose in codifications of the best interest duty accommodate consideration of employee interests as key stakeholders around AI adoption.

## 6. Employee Interests, AI Adoption and Four Doctrinal Corporate Purpose Models

In the common law world a trend over recent decades has been to transfer directors' duties including the best interests duty from case law to a statutory statement with a view to improving their accessibility (Ahern, 2012). This section presents a typology of four different models of statutory codification of directors' duties observed among countries that derive their company law model from pre-existing or existing Commonwealth membership with an associated lineage to British company law. The four models presented in ascending order of attentiveness to stakeholders are: shareholder primacy, enlightened shareholder value, stakeholder friendly and stakeholder value. In each category the interpretation and practical significance as regards employees around AI adoption by companies is probed. As discussed below, while shareholder primacy is linked with profit maximisation, enlightened shareholder value allows other stakeholders' interests to inform the decision on how to advance the interests of shareholders. Meanwhile a stakeholder-friendly model enables companies who so choose to widen their decision-making to consider interests beyond shareholders while a statutory stakeholder value model could potentially mandate appropriate consideration of stakeholder interests.

***6.1 Shareholder Primacy Models and Employee Interests***

Within a shareholder primacy model there need not always be goal conflict between shareholder goals and employee stakeholder goals. In some cases AI can help to ensure employee safety at work and reduce hazard in working conditions thereby augmenting employee welfare. Similarly, very long shifts and productivity expectations in companies may potentially lessen as AI can pick up the slack allowing for more leisure time. Integrating AI may therefore enhance employee welfare as well as meeting financial goals. Valuing employee perspectives around AI adoption enables better informed decisions to be made by directors, recognising the intricate connections between stakeholder interests and corporate success. Furthermore, meaningful employee engagement is likely to have tangible and intangible implications for the quality of internal and external corporate relationships and the company's reputation.

Common law traditional shareholder primacy jurisdictions such as Australia, Ireland and South Africa do not have any mention of stakeholder interests in their statutory formulations. As such, directors are not obliged in shareholder primacy models to consider employee interests, but equally they are not prohibited from doing so. The New Zealand Supreme Court gave credence to this expansive view in *Re Debut Homes Ltd (In Liq); Madsen-Ries v Cooper*,[22] ruling that section 131 of the Companies Act 1993 (prior to its subsequent amendment) gave carte blanche to directors to take into account considerations other than profit maximisation.[23] In practice there are many good reasons to do so from business to ethical to social. In shareholder primacy jurisdictions who do not include any mention of employee interests in their best interests duty, it is open to a board to choose to factor in stakeholder interests into decision-making (Baker, 2021). As ESG ("Economic, Social & Governance") values gradually gained a foothold in society and in company law, there has been an associated increasing recognition that in defining corporate purpose boards are free to set a broader set of values than profit maximisation (Lan & Wan, 2023). This is also the case in the United States subject to shareholder buy-in. In *Burwell v. Hobby Lobby Stores, Inc.,* 473 U.S. 682 (2014) the US Supreme Court stated:

> "While it is certainly true that a central objective of for-profit corporations is to make money, modern corporate law does not require for-profit corporations to pursue profit at the expense of everything else, and many do not do so."[24]

[22] [2020] NZSC 100.

On that basis, in shareholder primacy jurisdictions directors may (and arguably should) consider the impact of AI adoption on the current and future workforce, at a minimum where these considerations would matter to the directors' view of the best interests of the company (Glazebrook, 2019).

[23] However, the case centred on the interests of creditors and the observations on the interests of the company beyond that were strictly *obiter*; the Supreme Court declined to adjudicate between the enumerated models being (i) shareholder primacy (ii) inclusion of regard to stakeholder interests alongside profit maximisation and (iii) an entity-based approach that could potentially define the company's interests more broadly than traditional shareholder primacy.

[24] *Burwell v. Hobby Lobby Stores, Inc.,* 573 U.S. 682, 711-12 (2014).

### *6.2 The Enlightened Shareholder Value Model*

The picture for AI adoption and employees does not change dramatically in the Enlightened Shareholder Value model seen in the United Kingdom. Section 172 of the Companies Act 2006 set a precedent, but stopped short of a full-blown stakeholder value approach (Keay & Iqbal, 2019). To comply with it, directors in promoting the success of the company have a duty to reflect on how best the company can sustainably generate wealth over the long-term to comply with their section 172 duty. This follows the shareholder primacy model. However, the twist is that they must nevertheless in doing so "have regard to" the interests of named constituencies including "the interests of the company's employees" (section 172(1(b)). This marks directors' cards that in making decisions around AI adoption they must act in the long term interests of shareholders while "having regard to" other constituencies including employees. As Cheffins notes, with Enlightened Shareholder Value's introduction in the UK, "[t]he explicit invocation to take into account considerations in addition to shareholders likely enhanced board discretion" (Cheffins & Williams, 2021, p.1608). On the other hand, in no sense is there an expectation that stakeholder interests override those of shareholders in the success of the company (Johnston, 2017, p. 1032; Talbot, 2016, p. 515). The Enlightened Shareholder Value model validates the relevance of employees as stakeholders but does not elevate corporate purposes to a stakeholder value model given that section 172 expressly signals that the directors must act in the long-term interests of the members of the company. Indeed Bebchuk and Tallarita view operationalising stakeholderism within enlightened shareholder value as equivalent in outcome to shareholder primacy given the inevitable trade-offs that would need to apply in practice where the dominant concern continues to be shareholder value (Bebchuk & Tallarita, 2020, Part II). As such they dub the promise of stakeholderism “illusory”.

Thus directors making tough choices around AI-related layoffs and redeployment of workers are likely to be legally protected by taking a long-term approach to the company’s survival into the long-term. Doctrinally, any regard being had to employee interests in making decisions around AI adoption is coloured through the overall mandate in section 172 of ensuring the company continues to thrive for the long-term term interests of its members. This cue has significance as AI adoption is often motivated by the desire to ensure that the company adapts to ensure it continues to succeed in a changing world. Given the woolliness of the phrase “have regard to”, the section 172 approach allows boards who choose to do so to take employee interests seriously or less seriously in estimating what would promote the success of the company in the long-term. Thus one can necessarily expect an unevenness as to how valued employees are in crucial strategic decisions around AI that impact on their interests. Keay raises the interesting point that “considering the interests of employees might enable [the company] to recruit more qualified workers and to retain those who are invaluable to the operations of the company” (Keay, 2014, para. 6.117). This suggests the legitimacy of directors’ consideration the suitability of existing and future employees to serve the changing needs of the company , rather than just focusing on the interests of existing non-AI adopters in role entrenchment and employment preservation. Obviously, where employees are being asked to devote their time to training AI tools to do their job, only to find that their success in doing so has put colleagues out of a job, this will not be morale-boosting.

### *6.3 Stakeholder-friendly Models*

Jurisdictions with stakeholder friendly models of corporate purpose have embraced the ability for boards to opt-in to fashioning their own vision around stakeholder relevance. Both Canada[25] and New Zealand[26] revisited their codification landscape to specifically enable (but not mandate) directors' freedom to incorporate more inclusive consideration of stakeholder interests such as employees. In deciding questions around AI adoption, stakeholder-friendly legislative formulations of the best interests duty in Canada and New Zealand enable (but do not compel) directors to take a rounded view that is not only focused on the bottom line, but could also take account of the implications going beyond shareholder interests to encompass a broader range of stakeholder interests. This could yield more nuanced and meaningful interactions at board level around AI adoption and employee interests or at least provide an anchor for boards who were already adopting a stakeholder-inclusive approach. This type of holistic approach to the welfare of all would also be consistent with the Ubuntu approach whereby 'everyone will prosper' (Pellgrini Van Rooyen, 2021, p.7485).

Potentially, where boards buy in with hearts as well as minds, the spirit of stakeholder value can be ignited in thoughtful deliberations and engagement with stakeholders. On the other hand, Canada and New Zealand's opt-in stakeholder friendly models are facilitatory rather than mandatory. This contrasts to than the UK Enlightened Shareholder Value model where directors must "have regard to the named stakeholders' interests" in the context of ascertaining what would advance the long-term interests of the shareholders. The stakeholder-friendly models seen in Canada and New Zealand are not stakeholder value-–there is a marked disparity between the ceiling level (full consideration of stakeholders) as compared with the floor level (non-mandatory ability to consider stakeholders). Thus boards' choices around respect for employee interests may differ widely depending on how much stress is placed on the stake of employees in the corporate endeavour. Given the optionality, these stakeholder friendly legislative amendments have been branded, at best, principled, at worst, "hashtag capitalism" (Kamalnath, 2024; Watson & Buckley, 2024).

[25] Canada Business Corporations Act, s.122 (1.1).

[26] Companies Act 1993, s.131(5) (from the Companies (Directors' Duties) Amendment Act 2023.

In a curiously swift political *volte face* it is now proposed to remove Aotearoa New Zealand's legislative amendment (Ministry of Business Innovation and Employment "Appendix 1: Proposals in the Companies Act, Limited Partnerships Act, and Insolvency Act with a Regulatory Impact Analysis Exemption" (31 July 2024)' ; Underwood. & Buckley, 2025). The motivation has not been overtly expressed, other than to affirm that the pre-existing text would have allowed incorporation of stakeholder interests so express mention of this is perhaps unnecessary. One could nonetheless read this as signalling a desire by the Government to reaffirm a shareholder primacy model or a desire not to overcomplicate the duty for directors. If it represents a shift away from stakeholder value this is unfortunate for employee welfare as it collides with the AI boom that is happening now.

***6.4 The Stakeholder Value Model and AI adoption***

Positioning itself at the apex of approaches that challenge a shareholder primacy model that is focused on shareholder gain in analysing AI adoption, a pluralist or fully stakeholder-driven approach to the best interest duty would allow for a deeper consideration of the trade-offs that AI adoption may entail. A pluralist model would reshape traditional approaches to the duty by enabling direct consideration of stakeholder interests concerning AI adoption such as those of employees, customers, the community, suppliers, and the environment. Presenting the corporation as a social actor, stakeholder theorists and business ethicists have been challenging the shareholder primacy thesis for a long time (Blair, 2019, Donaldson & Preston, 1995; Freeman, 1984). Section 166(2) of India's Companies Act, 2013 requires directors to act "in the best interests of the company, its employees, the shareholders, the community and for the protection of the environment." Unlike the United Kingdom's Enlightened Shareholder Value model, no hierarchy or balancing act is expressly laid down, making this tricky territory for boards to navigate with respect to shareholder and stakeholder interests (Ramya & Ramesh, 2024).

An interesting thought is whether in a stakeholder-inclusive model directors should only consider the interests of current employees. What if a long-term approach was used that required their interests in the long-term to be looked at? That could dramatically alter the lens through which decision-making around AI reshaping roles and processes is viewed. A further argument that could be extrapolated to consider the impact of AI for our society is the argument made for an intergenerational justice approach that would require the wellbeing of future generations to be prioritised (Cheong, 2025.). Instrumentalising this would be tricky.

Taking account of varied and potentially conflicting interests becomes more complicated the further one advances into models that expressly acknowledge the need to take account of multiple stakeholder interests. This resonates in the context of decision-making around AI. Stakeholder-friendly, Enlightened Shareholder Value and stakeholder value models are non-directive on how the issue of conflicting stakeholder interests should be handled, a matter which has always dogged stakeholder theory. This could have particular significance in a battle between the interests of employees as stakeholders and the de facto preferment of AI agents as low cost and potentially superior servants of the company. Within team production theory the board's role is "to mediate among important competing interests in the corporation" (Blair, 2019, p. 199). Doctrinally, the resolution of multiple stakeholder tension, as with the tension between stakeholder and shareholder interests, lies with the directors on a case by case basis. Reflecting that, the courts unsurprisingly have not evinced a willingness to allow shareholders in derivative actions to seek dictate how competing considerations should be weighed in the balance. That this is a commercial decision, a classic business judgment for directors, was made crystal clear in the bellwether *ClientEarth* litigation which sought to open up the perceived inadequacy of Shell plc's consideration of climate protection interests.[27] In line with this, the courts have declared themselves "ill-equipped" to engage in a weighing of the various considerations which has been judicially dubbed as "essentially a commercial decision".[28] These sentiments have been echoed in courts across the common law world, indicating an aversion to second-guessing directors' view of what is in the best interests of the company.[29] This does not bode well for any attempt to argue that employee interests should come first. However, the decision of the Supreme Court of Canada in *BCE Inc. v. 1976 Debenture Holders*[30] goes somewhat further in suggesting that where the interests of stakeholders conflict, the duty to act in the best interests of the corporation places an onus on directors to act fairly as between stakeholders "commensurate with the corporation's duties as a responsible corporate citizen",[31] while acknowledging that it will need to be a directorial assessment as to whose interests prevail should a conflict arise.

[27] *ClientEarth v Shell* [2023] EWHC 1897 (Ch).

[28] *Iesini v Westrip Holdings Ltd* [2009] EWHC 2526 (Ch). [85].

[29] See, for example, *Re Debut Homes Limited (in liq); Madsen-Ries v Cooper* [2020] NZSC 100.

## 7. Restricted Duty Enforcement Pathways for Employees as Stakeholders under Hard and Soft Law

### *7.1 Employees and derivative actions*

Employee rights relevant to changing work practices and dismissal are largely found in employment contracts and employment law precepts including those which prescribe procedure around redundancy. As regards the best interests duty, even though employee interests are sometimes name checked in legislative formulations, generally they are assigned no direct right of enforcement. Scholars have on occasion argued for giving employees standing to bring a derivative claim (Safari & Gelter, 2019). Some corporate law landscapes are already potentially open to this. South African company law permits trade unions and employee representatives to initiate derivative actions.[32] Meanwhile Canada's federal corporate law landscape is open to the court exercising its discretion to allow any "proper person to make an application."[33] In other less flexible jurisdictions, workers or representatives of workers could acquire shareholder status with a view to attempting to take a derivative action arguing that the lack of consideration of employee interests is inextricably tied to the welfare of the company.

[30] 2008 SCC 69; [2008] 3 SCR 560.

[31] Para. 82.

[32] Companies Act 2008 (South Africa), s.165(2).

[33] Canada Business Corporations Act 1985, s.238(d).

That said, this is a procedurally fraught path. Corporate law makes deliberate choices to protect boards from litigation and there is precious little prospect for derivative actions querying a board's corporate purpose values or business judgment to succeed (Ahern, 2011). Overall, the derivative action does not present a promising doctrinal pathway in the face of compelling corporate interest considerations. In the United Kingdom with its Enlightened Shareholder Value model, having regard to the strictures around pursuing derivative actions, the odds are stacked against these claims gaining traction at the permission stage.[34] Decisions made by UK boards around AI adoption that countenance or result in labour force displacement are likely to be safe from challenge provided the minutes of the board meeting indicate that the interests of employees were considered in even the most cursory fashion before making the decision. This demonstrates that, however formulated, the best interests duty in combination with the enforcement regime for directors' duties do not lead to strong protective norms for workers that can be readily enforced by them. It is therefore important to cast the net wider in searching for how employee interests can be looked after in strategic board decision-making around firm AI adoption.

***7.2 Soft law and employee engagement and voice***

[34] *ClientEarth v Shell* [2023] EWHC 1897 (Ch).

Soft law corporate governance requirements that sit atop hard law requirements function to support keeping normative values around employee interests and voice in the picture. Although not based in duty or hard law, corporate governance codes can potentially exert powerful norms (Hill, 2020). The UK Corporate Governance Code expects companies to have appropriate engagement with stakeholders and shareholders (Financial Reporting Council, 2024, Section 1, Principle D). In addition, it prompts that "[t]he board should ensure that workforce policies and practices are consistent with the company's values and support its long-term sustainable success." (Financial Reporting Council, 2024, Section 1, Principle E). Furthermore, it is expected that the workforce can initiate dialogue to raise any concerns (Financial Reporting Council, 2024, Section 1, Principle E). These provisions on their face acknowledge the role of employees as one of the stakeholders contributing to sustaining company success in the long-term, consistent with team production theory.

Unlike in some countries in continental Europe, for common law countries employee engagement remains fairly new territory (Kokkinis & Segakis, 2020). For companies within the purview of the UK Corporate Governance Code, employee-dialogic interaction is formally shaped through mandating specific mechanisms for employee voice. These include a designated employee executive director or Non-Executive Director on the board or a workforce advisory panel. It has, however, been cogently argued that stakeholder representation on boards is not likely to be effective, in part because it is likely to run into conflict with shareholder primacy goals (Bebchuk & Tallarita, 2020, pp. 159-161).

These soft law provisions do floodlight the view that employees are important stakeholders who matter to corporate success. Nonetheless, the crux is that companies can either pursue a race to the top in meaningfully developing employee voice mechanisms or engage in hollow, tokenistic efforts in line with the soft law nature of the Uk Corporate Governance Code. Scholars are thus somewhat dubious in relation to achievable outcomes and the soft law approach to employee representation has been branded an obvious "non-starter" (Villiers, 2025, p. 320). Under this school of thought, in the absence of a fundamental reform corporate purpose, an employee engagement expectation based in soft law is unlikely to appreciably assist employees who fear the impacts of AI adoption in having their voice genuinely heard (Rees & Briône, 2024).

In the gap, unions are being asked to step up to provide a voice for workers around the transformation of work and to contribute to policy formation around AI (De Stefano & Doellgast, 2023; Velazquez, 2024). Unions have a role to play in shaping companies' choices and it is prudent for boards to engage with them early and frequently. In the United Kingdom, the work of the Trade Union Congress led to a Private Member's Artificial Intelligence (Regulation and Workers' Rights) Bill in 2023 that, if it had proceeded, would have treated dismissals based on an unfair reliance on high-risk AI deployment as unfair.[35] The fact that this did not bear fruit does not negate how fraught an issue AI-related displacement is and well-advised boards should meaningfully engage with their workforce around the underlying sentiments and embrace the values of transparency, fairness and dignity. The bigger picture view is that maintaining good employee relations is definitely in the company's best interests, both for the company's benefit as well as for its own sake. Directors should heed the value of going beyond the law through good faith engagement with employees to preserve employee morale and affirm their dignity. This is in line with the spirit of the law in context approach advocated for below.

[35] The subsequent Artificial Intelligence (Regulation) Bill HL 76 introduced in the House of Lords on 4 March 2025 did not contain any such provision.

**8. The Case for a Law in Context Approach to Employees as Stakeholders**

In the four different models of corporate purpose/best interest duty examined above, directors are enabled to take account of employees' interests but not legally mandated by any given formulation of the best interest duty to allow impacts on employees to outweigh shareholder interests. Furthermore, as to the potential for complaints around any perceived perfunctoriness of attentiveness to and consideration of employee interests, directors might be inclined to dismiss the enforcement risk as negligible. Certainly the procedural hurdles to enforcement severely limit the effectiveness of the best interests duty as a vehicle of law in action to offer enforceable protection to employees around the AI-related concerns of employees. Similar concerns negate the effectiveness of supporting soft law provisions. Assessing the potential to address employee interests in directors' handling of AI matters therefore requires recognising that the sphere of influence is wider than the framing provided by the best interests duty and the soft law expectations of corporate governance codes around employee voice. Indeed, expectations of boards are shaped by dynamic, policy and public discourse in areas such as climate impacts (Sarra, 2025). Moreover, even with shareholder primacy, markets can act as a corrective where regulation does not including through "consumer aversion, and employee attrition" (Lipton, 2025, p.86).

Normatively, the expectation that directors be responsive to the likely impact on the company's workforce goes to directors' good faith when dealing with employees as the pre-eminent stakeholder class beyond shareholders. Consistent with this, governance is not only about minimum legal compliance (Hill, 2025), and as a matter of principle and fair dealing that goes to corporate reputation, it would be foolish for boards not to consider how to engage with employees around AI as early as possible and as regularly as possible as AI roll-out plans are developed. If a company has chosen to foreground itself as 'sustainable' and as providing a worker-friendly environment, negative impacts on the workforce including AI-related layoffs may publicly butt up against these values and attract adverse commentary.

In advocating for a law in context approach, the contention is that despite being well-insulated from corporate purpose suits, directors ought nevertheless to choose to consider and give due weight to stakeholder considerations when making strategic decisions around AI adoption in the company's operations. Law in context/law in society scholars emphasise the importance of contextualising how legal norms interface with societal, moral and cultural norms (Selznick, 2003). These in turn influence legal norm building. A law in context approach is particularly suited to situating directors' duties in the context of the rapidly evolving landscape around AI adoption, where typically legal norms trail societal change. Thus it is vital for directors to reference commercial developments and societal controversies precisely because legal norms are "conditioned by culture and social organization" (Selznick, 2003, p.177). In short, law must ultimately yield to societal context on the ground as norm building around AI adoption and its implications are evolving. The legal order is coloured by the societal context and expectations including those driven forward by activists and scholars. Indeed, one suggestion is that executive remuneration could be tied to achievement of employee interest outcomes (Salazar, 2023).

Employee morale is intangible but its loss is potentially devasting for corporate performance and reputation. Boards should therefore deeply consider employee interests in their deliberations around AI. Directors need to be conscious of employee concerns around how AI in the workforce impacts on them including concerns around the dehumanisation of the workplace (Atkinson, 2021). Even the threat of potential technological change can create distrust and lower employee morale due to fears of job displacement. More seriously, poorly handled AI-related redundancies may lead to cost efficiencies lauded by investors and satisfying shareholder primacy goals, but overall company employee morale may suffer with resulting negative impacts on employee retention, productivity and employee relations. This includes the replacement of management discretion with AI driven selection of workers for termination/redundancy and promotion (Crist, 2025). This points up the need for directors to approach AI adoption in a way that goes beyond a minimum compliance mentality to the best interests duty as regards factoring in employee interests. Directors may adopt a deteriorating approach to valuing employees as stakeholders where their value to the corporation is seen as disposable in the face of increasing reliance on AI task substitutes. This may not warrant a derivative action for breach of the best interests duty. However, it could lend itself to a public victory played out in the court of public opinion, giving rise to a legitimacy and image problem for the board and the company if it is perceived to have treated workers badly. This fits with Galanis' concept of "social logic" as providing a feedback loop in spite of corporate law's focus on wealth accumulation (Galanis, 2021, p.24). He argues that where corporate law blocks responsiveness to social change, it becomes a "socially hazardous institution" (Galanis, 2021, p.15). Pondering this, perhaps reflective managers will realise that a focus on profit-first decisions without considering the broader implications which AI adoption presents for employees and customers is likely to be a fool's game if negative social feedback is the result of an unbalanced concentration on using AI to drive corporate efficiency while downplaying the role of soft skills and the human touch. Companies who fail to grasp this lesson may well be within the letter of the law of the best interests' duty but they may fall spectacularly foul of employee, customer and public opinion.

Rather than pursuing a tick box approach to employees as stakeholders, if directors want to meaningfully cultivate a fairness approach to the interests of employees, then boards should be ensuring that there is a meaningful, open dialogue. This should include creative exploration of opportunities for upskilling and retraining employees rather than simply moving people to selection for redundancy. Going beyond the law would mean the board actively engaging with the context that, as one study suggests, between 25 and 40 percent of occupations are "AI retrainable" (Hyman et al., 2025). Companies can choose to provide opportunities for employee reskilling for employees to adapt whose work would be significantly impacted or made redundant by AI (Morandini et al., 2023; Seelent et al., 2025; Spitko, 2024). Thus a fairness and inclusivity approach to employees interests would involve the board first taking corporate responsibility for AI-related role displacement, and second, acting in the interests of employees in their careers in a positive way that simply making employees redundant would not. An awareness of reskilling initiatives would benefit wider employee morale and firm reputation as challenging organisational change and transformation is managed.

Not every AI adoption scenario will lead to employee retention or reskilling. That being so respect for the dignity of employees should inform communications and boards should be mindful of "tone at the top". Failing to appropriately consider employee dignity may deliver a powerful negative reputation ripple magnified by social media. Meaningful and caring engagement with affected employees matters to employees, but also to the wider community and customer base, and thus to the company's goodwill and reputation as a good faith actor. Two examples illustrate this. Accenture CEO Julie Sweet, recognising the criticality of advanced AI to the future, announced that the firm would be "investing in upskilling our reinventors" while in rather brusque corporate speak, she indicated that workers for whom reskilling was not viable would be "exiting on a compression timeline" (Bhaimiya, 2025; Foley, 2025). The expression of the message here was deeply unfortunate, suggestive of a disposable approach to employment that lacked empathy with the workers who it was planned to rapidly lay off. In another case, the Commonwealth Bank of Australia was forced to reverse a plan to fire 45 call centre workers that it intended to replace with an AI chatbot including a long-standing employee who had spent years helping to train the AI chatbot to do the job. The bank acknowledged that mistakes had been made around respect and support (Perrie, 2025). Some companies are going so far as to engage in a disingenuous practice of deliberately understating the extent of the company's AI adoption and investment to avoid transparency that could lead to public and stakeholder scrutiny of social and environmental implications including labour force reduction, a phenomenon known as "AI-masking" (Donelson et al., 2025).

Cases such as these illustrate that directors would do well to go beyond base level compliance with the duty to act in the company's interests and act in line with the concept of governance beyond compliance to meaningfully acknowledge employees as stakeholders with dignity and to engage with their role and process needs. After all, directors taking steps to engage with and to understand the viewpoints of different classes of employees, along with appropriate handling of a diverse range of perspectives is appropriate to companies being not just economic actors, but societal actors in the age of AI.

## 9. Conclusion

Attentiveness to the significance of AI for a company's business constitutes a pressing matter of strategy for directors under their best interests duty with significance for a company's competitiveness and to its profitability in traditional shareholder primacy terms. As such, it is appropriate for directors to be alive to the value proposition that AI may present for their company as well as to how failure to integrate AI into the company's operations may compromise the company's market positioning. Widening out the focus, this article has explored the nuance that is added to the mix by the status of employees as stakeholders and their needs around AI adoption, having concluded that we are not yet at the point where AI actors join or supersede human workers as stakeholders in the corporation. From the four types of best interests duty formulation surveyed, it can be concluded that both races to the top and races to the bottom remain entirely possible for directors across the spectrum of corporate purpose models. Consideration of employee interests around AI adoption is permissible in shareholder primacy and stakeholder-friendly corporate purpose models. Meanwhile an enlightened shareholder value requires this while a stakeholder value model would encourage weight being given to employees' concerns. Nonetheless, within the four doctrinal models of corporate purpose examined, however enabling they may seem, none of the typologies of corporate purpose formulations are stakeholder-first or employee-driven regimes (with the possible exception of the stakeholder model seen in India which overtly presents an equality of esteem approach).[36] Thus a race to the top could see employee boardroom representation and meaningful employee consultation and engagement while a race to the bottom would see minimum compliance with no heed or lip service to employees as stakeholders in directors' decision-making around AI adoption and associated workforce restructuring. As such, scope for directors to single-mindedly pursue a shareholder-focused, profit maximisation agenda around AI as discussed in Section 5 is alive and well. In that mindset employee interests are negatively associated with increased labour costs and reduced profitability and directors may choose to leave employee protection to the preserve of contract and employment law.

Nonetheless, what this article has highlighted is that leaving employees out in the cold would be a shortsighted approach given the nuances that are beginning to emerge in relation to how a responsible company should respond to its employees on its journey of AI integration. Setting the analysis of employee interests and the best interests duty within a broader law in context setting fits within the premise that the norms that constrain directors are multi-headed (Hill, 2020). Much of the ballast that should influence AI-related decision-making by directors is supplied outside of courtrooms by emerging best practice, societal demands, and external pressures. Therefore this article has advocated the merits of directors going beyond mere compliance with the letter of the best interest duty in favour of a far more holistic approach to understanding and meeting the interests of the company. This involves navigating appropriate engagement with and treatment of employees. Normatively, a law in context approach recognises that the substance of the best interests duty in terms of what is expected of directors ought to be informed by the wider commercial, social and legal landscape rather than simply by doctrine. Employee morale and public reputation consequences accrue from failing to conform to law in context ethical expectations of the corporation as a social actor. While judicial scrutiny is rare, public scrutiny of companies' decision-making around employees and AI is beginning to reveal clashes with emerging social and fairness norms, demonstrating the potential for negative reputational impact for companies. This bolsters the need for a law in context approach to the interests of employees that goes beyond a minimum compliance approach. It is thus vital for directors and their legal advisors to remember that how they handle AI adoption can cause reputational damage for companies and their directors irrespective of whether liability for breach of the best interests duty can be driven home. Opportunities for employee reskilling and retraining mean that this need not be the great AI goodbye. For their part, as stakeholders in the corporate endeavour, employees need to be flexible, to meet boards halfway by exhibiting openness and adaptiveness to retraining and role pivoting.

---

[36] The equality of esteem is in practice detracted from by the non-accessibility of the derivative claim to non-shareholder stakeholders.